\documentclass[showpacs,showkeys,preprint]{revtex4}

\def\ii{\'\i}

\begin{document}

\title{On the speed of pulled fronts with a cutoff}
\author{R. D. Benguria and M. C. Depassier}
\affiliation{
 Facultad de F\'\i sica\\
     Pontificia Universidad Cat\'olica de Chile\\
           Casilla 306, Santiago 22, Chile}

\date{\today}

\begin{abstract}
We study the effect of a small cutoff $\epsilon$ on the velocity of a pulled front in one dimension by means of a variational principle. We obtain a lower bound on the speed dependent on the cutoff, and for which the two leading order terms  correspond to the Brunet Derrida expression. To do so we cast a known variational principle for the speed of propagation of fronts in new variables which makes it more suitable for applications.
\end{abstract}

\pacs{47.20.Ky,05.45.-a,05.70.Ln, 02.30.Xx}
\keywords{reaction-diffusion equations,cut-off,traveling waves,critical wave speeds, variational principles}

\maketitle

\section{Introduction}
\label{intro}
In several problems arising in physics, population dynamics, chemistry and other fields, it is found that a small perturbation to an unstable state leads to a propagating front joining the unstable to an stable state. The simplest model of such phenomenon is provided by the scalar reaction diffusion equation
\[u_t = u_{xx} + f(u)\]
where the reaction term $f(u)$ is a nonlinear function with at least two fixed points, one stable and one unstable.
Without loss of generality we assume that there is an unstable fixed point at $u=0$ and a stable fixed point at $u=1$. The reaction term $f(u)$ obeys additional requirements depending on the phenomenon under study. In the present work we shall be interested in two generic classes. The first class, which we label type A, is that for which $f>0$ in $(0,1)$, the second class, type B, also called the combustion case, is that for which $f=0$ in $(0,a)$, and $f>0$ in $(a,1)$. It was proved by Aronson and Weinberger \cite{AW78} that sufficiently localized initial conditions evolve into a monotonic front joining the stable to the unstable state. In case B there is a unique speed for which a monotonic front exists. In case A,  the front propagates with the minimal speed for which monotonic fronts exist.   This minimal speed satisfies
\begin{equation}
2 \sqrt{f'(0)} \leq c^* < 2 \sqrt{\sup_{0 \le u \le 1} (f(u)/u)},
\end{equation}
result also found by Kolmogorov, Petrovsky and Piskunov (KPP) \cite{KPP}. For the classical Fisher--Kolmogorov \cite{Fisher,KPP} equation \[u_t = u_{xx} + u(1-u),\]
the upper and lower bounds coincide and the speed is exactly the so called linear or KPP value $c_{KPP}=2\sqrt{f'(0)}$. Fronts for which this is  the minimal speed  are called pulled since this is the speed obtained from linear considerations at the leading edge of the front.  In all cases the speed can be calculated from the integral variational principle \cite{BD96c}
\begin{equation}
c^2 = \sup_{g(u)} \,2 \,\frac
{ \int_0^1 f(u) g(u) d\,u}
{\int_0^1 g^2(u)/h(u) d\,u}
\label{vp1}
 \end{equation}
where the supremum is taken over all positive monotonic decreasing functions $g(u)$ for which the integrals exist and where $h(u) = -g'(u)$. Moreover, the supremum is always attained for reaction terms of type B and for reaction terms of type A it is attained whenever $c> c_{KPP}$.

Reaction diffusion equations of type A are often used to model phenomena in population dynamics, with the assumption that the number of particles or individuals is large. It was noticed by Brunet and Derrida \cite{brunet-derrida} that the effect of a finite number of particles can be   modeled by reaction terms of type A with a cutoff $\epsilon = 1/N$, where $N$ is the average number of particles at the saturation state of the front. The effect of a cutoff on the fronts was studied for the case
$
f(u) = u - u^3$ and it was found that the speed of the front with a cutoff is given approximately by
$$
c \approx 2 -\frac{\pi^2}{(\log \epsilon)^2}.
$$
This result was obtained by a matching approach. Recently this result has been  proven rigorously by a geometric method \cite{Kaper}.
The precise dependence of the speed of the front on the cutoff is not universal. An example of this non universality was shown by introducing a small region of vanishing slope next to the cutoff \cite{panja1}, the speed in this case turns out to be larger than the KPP value.
The effect of a cutoff on reaction diffusion equations is of relevance not only as a model of populations with a large but finite number of individuals, it is also relevant to the study of noisy fronts \cite{conlon} and to some problems of particle physics. See for example \cite{ brunet-mueller} and references therein.

The purpose of the present work is to show how the speed of a pulled front with a cutoff can be found from the variational principle (\ref{vp1}). It is important to notice that the effect of a cutoff of a reaction term of type A is to transform it into a reaction term of type B, reaction terms for which the supremum in (\ref{vp1}) is attained and for which a unique speed exists.

To obtain this result we reformulate the variational principle in a new way better suited to treat the fronts with cutoff. We apply this new form of the variational principle to the reaction term considered by Brunet and Derrida, but the results are of more general validity. We find a lower bound on the speed which depends  on the cutoff, for which the leading order is the Brunet-Derrida expression. That is, we show that,
\[
c \geq c(\epsilon) \approx 2 - \frac{\pi^2}{(\log \epsilon)^2} + \rm{h.o.t}
\]
The same result can be obtained by using the alternative variational principle for the speed \cite{BD96a}.

\section{A simpler form for the variational principle}

In this section we introduce new variables which render the variational formula (\ref{vp1}) for the speed simpler to apply, particularly to the case of fronts with a cutoff. As an application of this simpler form we show 
in the Appendix A how the linear or KPP value $c_{KPP}=2$ is obtained for the Fisher Kolmogorov equation.

The variational expression (\ref{vp1}) implies that for any admissible trial function $g(u)$,
\[
c^2 \ge \,2 \,\frac
{ \int_0^1 f(u) g(u) d\,u}
{\int_0^1 g^2(u)/h(u) d\,u}
\]

It was shown in \cite{BD96c} that the trial function $\hat g(u)$ for which equality holds diverges at $u=0$, so it is convenient to consider trial functions  which in addition to the requirements $g(u) > 0$, $g'(u) < 0$ also satisfy $g(0)\rightarrow \infty$.

Since $g(u)$ is a monotonic decreasing we may perform the change of variables
$$
u = u(s), \qquad {\rm where} \qquad s= 1/g,
$$
and consider $s$ as the independent variable in (\ref{vp1}). With this change of variables we find
$$
\int_0^1 f(u) g(u) d\,u= \frac{F(1)}{s_0} + \int_0^{s_0} \frac{F(u(s))}{s^2} d\,s,
$$
where $s_0= 1/g(u=1)$ is an arbitrary parameter and
$$
F(u) = \int_0^u f(q) dq.
$$
The denominator becomes
$$
\int_0^1 \frac{g^2(u)}{h(u)} d\,u = \int_0^{s_0} \left( \frac{d u}{d s}\right)^2 d\,s.
$$
In this new variables the variational principle becomes
\begin{equation}
c^2 = \sup_{u(s)} 2 \frac{ F(1)/s_0 + \int_0^{s_0} F(u(s))/s^2 d\,s}{\int_0^{s_0} \left( d u /d s\right)^2 d\,s},
\label{newvp}
\end{equation}
where the supremum is taken over positive increasing functions $u(s)$ such that $u(0)=0$, and for which all the integrals in (\ref{newvp}) are finite.

In the Appendix A, we illustrate how to use this variational principle to show that for profiles satisfying 
the KPP criterion (i.e., $f(u) \le  f'(0) u$, for all $0 \le u \le 1$), $c=2 \sqrt{f'(0)}$.

\section{The speed of the front with a cutoff}

In this section we consider the speed of a front for a reaction term with a cutoff. Even though we choose a specific reaction term the results obtained are valid for a larger class of reaction terms. We choose the same reaction term studied previously by Brunet and Derrida, namely
$$
f(u) = \left\{ \begin{array}{ll}
              0
             & \mbox{if $0\leq u \leq \epsilon$} \\
             u - u^3 &\mbox{if $\epsilon < u < 1.$}
           \end{array} \right.
$$
For this reaction term
$$
F(u) = \left\{ \begin{array}{ll}
              0
             & \mbox{if $0\leq u \leq \epsilon$} \\
             u^2/2 - u^4/4 - \epsilon^2/2 + \epsilon^4/4 &\mbox{if $\epsilon < u < 1$}
           \end{array} \right.
$$
so that $F(1) = 1/4 - \epsilon^2/2 + \epsilon^4/4.$
We will show that for a certain trial function the variational formula (\ref{newvp}) yields
$$
c^2 \geq c^2(\epsilon) \approx 4 \left( 1 - \frac{\pi^2}{|\ln \epsilon|^2} +\ldots\right).
$$
For the sake of clarity we postpone  until section 4 the construction of the trial function.

Choose the trial function
\begin{equation}
u(s) = \left\{ \begin{array}{ll}
              s
             & \mbox{if $0\leq s \leq \epsilon$} \\
             A \sqrt{s} \cos \phi(s) &\mbox{if $\epsilon < s< s_0$}
           \end{array} \right.
\label{trial}
\end{equation}
where
\begin{equation}
 s_0 = 1/\epsilon, \qquad \qquad  \phi(s) = \omega \ln (s/\epsilon) - \phi_*,
\end{equation}

\begin{equation}
 \omega = \frac{\phi_*}{|\ln \epsilon|}, \qquad \qquad  A = \sqrt{\epsilon}\left( 1+ \frac{1}{4\omega^2}\right)^{1/2},
\label{aw}
\end{equation}
and where $\phi_*$ is the solution of
\begin{equation}
\phi_* \tan \phi_* = \frac{1}{2} |\ln \epsilon|.
\label{valor}
\end{equation}
Notice that these definitions imply that in the range $\epsilon < s < s_0$, $-\phi_* < \phi(s) < \phi_*$,
and that, for small $\epsilon$,  $\phi_* < \pi/2$. Therefore $u(s)$ is positive and monotonic increasing.

Having chosen a trial function it is straightforward to obtain a lower bound. The  numerator in (\ref{newvp}) is
given by
\begin{eqnarray}
N(\epsilon) &=& F(1)/s_0 + \int_0^{s_0} F(u(s))/s^2 d\,s 
\nonumber
\\
&=& \epsilon F(1) + \left(\epsilon -\frac{1}{\epsilon}\right)\left( \frac{\epsilon^2}{2} - \frac{\epsilon^4}{4}\right) + \frac{A^2}{4 \omega}(2 \phi_* + \sin 2\phi_*) - \frac{1}{4}\int_\epsilon^{1/\epsilon} \frac{u^4}{s^2} \, ds 
\nonumber
\\
&\geq&
\epsilon F(1) + \left(\epsilon -\frac{1}{\epsilon}\right)\left( \frac{\epsilon^2}{2} - \frac{\epsilon^4}{4}\right) + \frac{A^2}{4 \omega}(2 \phi_* + \sin 2\phi_*) + O(\epsilon^{3/2} (\ln \epsilon)^4).
\label{N1}
\end{eqnarray}
(See equation (\ref{B5}) in the Appendix B for the details on the estimation of the last integral in (\ref{N1})).
The denominator is given by
\begin{eqnarray}
D(\epsilon) &=& \int_0^{s_0} \left( d u /d s\right)^2 d\,s 
\nonumber
\\
&=& \epsilon + \frac{A^2}{8 \omega} \left[(1 + 4 \omega^2) 2 \phi_* + (1 - 4 \omega^2) \sin 2 \phi_*\right].
\label{D1}
\end{eqnarray}
We know then that
\begin{equation}
c^2 \geq c^2(\epsilon) = 2 \frac{N(\epsilon)}{D(\epsilon)}.
\label{nd}
\end{equation}
The bound above is rigorous and it is explicitly dependent on $\epsilon$.  Expanding $c(\epsilon)$ for small $\epsilon$ we obtain in leading order the desired result, (see part ii of Appendix B, in particular the derivation of (\ref{B8}))
\begin{equation}
c^2(\epsilon) = 4 \left( 1 - \frac{\pi^2}{|\ln \epsilon|^2}+ h.o.t.\right).
\label{bdr}
\end{equation}

It is not difficult to obtain higher order terms in the expansion of $c(\epsilon)$ but it is of no interest here.

\section{The trial function}

To find the trial function for which the maximum in the variational formula for the speed is atained we should solve the associated Euler-Lagrange equation. This is not possible in general since there are few exactly solvable cases. In the present situation we are interested in the effect of the cutoff on pulled fronts, that is, on fronts whose speed is determined from linearization at the leading edge, therefore we solve the Euler-Lagrange equation in the linear approximation.

The Euler-Lagrange equation for the variational principle (\ref{newvp}) is
$$
\frac{d^2 u}{d s^2} + \lambda \frac{f(u)}{s^2} =0,
$$
where $\lambda$ is a Lagrange multiplier. Even though it is unrelated to the present discussion, it is worth mentioning that this equation can be obtained by performing the change of variable $s =$ exp$(- c z)$ in  the ordinary differential equation $u_{zz} + c u_z + f(u)=0$, hence we identify the Lagrange multiplier with $1/c^2$.

First we obtain the adequate trial function for pulled fronts without a cutoff. In the linear approximation the Euler-Lagrange equation is
$$
\frac{d^2 u}{d s^2} + \lambda \frac{u}{s^2} =0,
$$
subject to $u(0) = 0$. The solution is of the form $u = s^{\alpha}$ where $\alpha$ is given by
$$
\alpha = \frac{1}{2} \pm \sqrt{1- 4\lambda}.
$$
As shown in Appendix A, the best bound is obtained for $\alpha\rightarrow 1/2$ hence the Lagrange multiplier in that limit is $\lambda= 1/4$.

 Next we construct the appropriate trial function for a pulled front with a cutoff. We must solve
 \[
 \frac{d^2 u_1}{d s^2} = 0 \qquad\qquad u_1(0) =0 \qquad \qquad {\rm for} \qquad 0<u< \epsilon.
 \]
 Since the variational formula (\ref{newvp}) is invariant to scaling in $s$ we may choose, without loss of generality,
$ u_1 = s$. Moreover, since this is valid for $0<u<\epsilon$,  we conclude that in this region $0<s<\epsilon$. To sum up we have
$$
u_1(s) = s, \qquad\qquad {\rm if} \qquad 0<s\leq \epsilon.
$$
For $u>\epsilon$ ($s>\epsilon$) but still small enough for the linear regime to be valid, we must solve
$$
 \frac{d^2 u_2}{d s^2} + \lambda \frac{u_2} {s^2} =0,\qquad \qquad {\rm with} \qquad u_2(\epsilon)= u_1(\epsilon), \qquad u'_2(\epsilon)= u'_1(\epsilon)
$$
The solution to this equation is straightforward, it is given by
$$
u_2(s) = A \sqrt{s} \cos \phi(s)
$$
where $$\phi(s) = \omega \ln\frac{s}{\epsilon} + \delta, \qquad \qquad \omega= \frac{1}{2} \sqrt{4 \lambda-1}.
$$
The constants $A$ and $\delta$ are found matching the solution to $u_1$ as indicated above.
Applying the matching conditions we obtain
\begin{equation}
A = \sqrt{\epsilon}\left( 1+ \frac{1}{4\omega^2}\right)^{1/2}, \qquad \delta = \arctan\frac{-1}{2\omega}.
\label{A}
\end{equation}
We must also require that $u(s)$ be positive and monotonic increasing. The first condition implies
$ -\pi/2 < \phi(s) < \pi/2$. The second condition, that $u(s)$ be monotonic increasing implies
$$
\tan\phi(s) \leq \frac{1}{2\omega}.
$$
Since $\phi(s)$ is an increasing function of $s$ we know that
$$
\arctan\frac{-1}{2\omega}=\phi(\epsilon) \leq  \phi(s) < \phi(s_0) \leq \arctan \frac{1}{2\omega}.
$$
It is intuitively evident that the best bound will be obtained when the maximum range for $\phi$ is allowed. We choose then
\begin{equation}
\phi(s_0) \equiv \phi_* = \arctan\frac{1}{2\omega}.
\label{fistar}
\end{equation}
With this choice, $\delta = \phi(\epsilon)=-\phi_*$. The only free parameter left is the arbitrary parameter $s_0$. To fix $s_0$ we observe that as $s\rightarrow s_0$ the solution must approach $u=1$. Since for pulled fronts without a cutoff $\omega=0$, we expect that for small $\epsilon$, $\omega$ will be small. Then from (\ref{fistar}) it follows that
$$
\phi_* = \frac{\pi}{2} - 2 \omega + h.o.t.,$$
hence
$$
u(s_0) = A \sqrt{s_0} \cos\phi_* \approx A \sqrt{s_0} \sin (2\omega) \approx 2 \omega A \sqrt{s_0} .
$$
From (\ref{A}) we see that for small $\omega$,
$$
A \approx \frac{\sqrt{\epsilon}}{2\omega}.
$$
Therefore, for small $\omega,$ $u(s_0) \approx \sqrt{\epsilon s_0}$. Requiring that $u(s_0) \rightarrow 1$ implies then $s_0 = 1/\epsilon$. With this choice for $s_0$ it follows that
$$
\phi(s_0) = \phi_* = - 2 \omega \ln\epsilon - \phi_*,
$$
hence
$$
\omega = \frac{\phi_*}{|\ln \epsilon|}.
$$
Replacing this value of $\omega$ in (\ref{fistar}) we obtain (\ref{valor}), with which the construction of the trial function is complete.

\section{Conclusion}

The purpose of this work was to study the effect of a cutoff on the speed of pulled fronts making use of the variational formulation for the speed. To do so we have rewritten the variational principle in new variables which
simplify the problem. An additional advantage of this reformulation is that the Euler-Lagrange equation for the maximizer is seen easily to be the equation of the traveling front.

Reaction terms with a cutoff belong to the class of general reaction terms for which a maximizer always exists and for which the speed is unique. If the original front without a cutoff is a pulled front then, with a cutoff, it is possible to solve the Euler-Lagrange equation in the linear approximation and obtain an upper bound on the speed. This value obtained from the linear equation is vaid only for sufficiently small cutoffs. The lower bound on the speed is a complicated function of the cutoff, the first two terms in the series expansion of this bound correspond to the approximate formula found by other approximate means. Here we have obtained not only the first two terms in the expansion but a rigorous bound on the speed.

It has been shown that small perturbations of the reaction term close to the cutoff have an important effect, and that the Brunet-Derrida term is not universal for all fronts with a cutoff. This can be expected  since the Euler-Lagrange equation in the linear regime  will be different in each case. A detailed analysis of this situation will be reported elsewhere.

In the present work we have studied the effect of a cutoff on a pulled front, the effect of a cutoff on pushed fronts
and bistable fronts has received less attention. A specific example is studied in \cite{kns}. General bounds on the speed have been obtained making use of the variational principle (\ref{vp1}) and exact solutions have been constructed for piecewise continuous functions \cite{mendez} .  These and other related problems will be the subject of future work.

\section*{Acknowledgements}
We acknowledge partial support of Fondecyt (CHILE) projects 106--0627 and 106--0651, and  CONICYT/PBCT Proyecto Anillo de Investigaci\'on en Ciencia y Tecnolog\ii a ACT30/2006.

\appendix
\section{}

Here we show how to recover the value $c=2 \sqrt{f'(0)}$ from the variational formulation (\ref{newvp}) 
for the speed of propagation of fronts when the profile $f(u)$ satisfies the KPP condition, i.e., when
$f(u) \le f'(0) u$, for all $0 \le u \le 1$. Let us denote
\begin{equation}
N = \frac{F(1)}{s_0} + \int_0^{s_0} F(u(s)) \frac{1}{s^2} \, ds,
\label{A1}
\end{equation}
and
\begin{equation}
D=\int_0^{s_0} \left( \frac{du}{ds} \right)^2 \, ds.
\label{A2}
\end{equation}
For the KPP case, $F(u) \le f'(0) u^2 /2$, hence $F(1) \le f'(0)/2$. Then, it follows from (\ref{A1}) that
\begin{equation}
N \le  M \equiv \frac{f'(0)}{2 s_0} + \int_0^{s_0} f'(0) \frac{u^2}{2 \, s^2} \, ds.
\label{A3}
\end{equation}
Intregrating the last term by parts and noticing that $u(0)=0$, $u(s_0)=1$, and that $\lim_{s \to 0} u/s = u'(0)$ exists, we get,
$$
\int_0^{s_0} f'(0) \frac{u^2}{2 \, s^2} \, ds
=\int_0^{s_0} f'(0) \frac{u^2}{2} \left(-\frac{d}{ds} \frac{1}{s} \right)  \, ds = -\frac{f'(0)}{2 s_0} + \int_0^{s_0} f'(0) u u'\frac{1}{s} \, ds.
$$ 
Therefore,
\begin{equation}
M^2 \le \left(\int_0^{s_0} f'(0) u u'\frac{1}{s} \, ds\right)^2  \le f'(0)^2 \int_0^{s_0} \frac{u^2}{s^2} \,ds
\int_0^{s_0} (u'(s))^2\,ds,
\label{A4}
\end{equation}
by Schwarz inequality. However, from (\ref{A3}) we have 
$f'(0) \int_0^{s_0} u^2/(2 s^2) \, ds \le M$, and inserting this in (\ref{A4}) we finally get,
\begin{equation}
M \le 2 \, f'(0) \int_0^{s_0} (u'(s))^2 \, ds.
\label{A5}
\end{equation}
Now, from (\ref{A2}), (\ref{A3}), and (\ref{A5}) we have that
\begin{equation}
2 \frac{N}{D} \le 4 \, f'(0),
\end{equation}
for all possible trial functions $u$. Therefore, taking the supremum of $2 N/D$ over all $u$, 
using (\ref{newvp}) we finally get,
\begin{equation}
c \le 2 \sqrt{f'(0)}.
\label{A6}
\end{equation}

On the other hand, choosing an appropriate maximizing sequence of functions, using the variational principle 
(\ref{newvp}),
we may show that $c \ge 2 \sqrt{f'(0)}$ and thus we  can conclude that $c= 2 \sqrt{f'(0)}$ in the KPP case.
 For that purpose, just consider the 
family of trial functions $u_{\alpha} = s^{\alpha}$, which are appropiate trial functions as long as 
$\alpha > \frac{1}{2}$. Evaluating the right side of (\ref{newvp}) with $u=u_{\alpha}$ and letting $\alpha \to 1/2$,
we get $c^2 \ge 2 \sqrt{f'(0)}$, which combined with (\ref{A6}) yields the desired result. Just to illustrate
this procedure consider the reaction term $f(u) = u - u^3$. Effectively, with this trial function (\ref{newvp}) implies
$$
c^2 \geq \frac{2}{\alpha^2} \left[ \frac{2\alpha-1}{4 s_0^{2\alpha}} +\frac{1}{2} - \frac{2\alpha-1}{4(4\alpha-1)} s_0^{2\alpha}\right].
$$
In the limit $\alpha \rightarrow 1/2$ we obtain $c^2 \ge 4$.

\bigskip

\section{} 
 
In this Appendix we show some details on how to get the lower bound (\ref{bdr}).

\bigskip
\noindent 
i) {\bf Estimating the integral $I \equiv \int_{\epsilon}^{s_0} u^4/s^2\, ds$ for the trial function $u$ given by (\ref{trial})}.
In order to estimate this integral we divide it 
into two parts, as follows,
\begin{equation}
I_1 = \int_{\epsilon}^{\epsilon^{-1/2}} \frac{u^4}{s^2} \, ds,
\label{B1}
\end{equation}and, 
\begin{equation}
I_2= \int_{\epsilon^{-1/2}}^{\epsilon^{-1}} \frac{u^4}{s^2} \, ds.
\label{B2}
\end{equation}
To estimate (\ref{B1}) we insert (\ref{trial}) and we get
\begin{equation}
I_1 = \int_{\epsilon}^{\epsilon^{-1/2}} A^4 \cos^4 \phi(s) \, ds \le A^4 (\epsilon^{-1/2} - \epsilon).
\label{B3}
\end{equation}
As for the second integral we use that $u \le 1$ to get
\begin{equation}
I_2 \le \sqrt{\epsilon} - \epsilon.
\label{B4}
\end{equation} 
Adding up this two integrals and using (\ref{aw}) we see that $I$ can be estimated from above by a term of 
order $\epsilon^{3/2} |\ln \epsilon|^4$, i.e.,
\begin{equation}
\int_{\epsilon}^{1/\epsilon} \frac{u^4}{s^2}  \, ds \le O( \epsilon^{3/2} |\ln \epsilon|^4),
\label{B5}
\end{equation}
which is small compared with 
$A^2/\omega \approx \epsilon/(4 \omega^3) = O(\epsilon (\ln \epsilon)^3)$,
since $\epsilon^{1/2} (\ln \epsilon) \to 0$ as $\epsilon \to 0$.

We have shown that the contribution of the nonlinear term $I$ can be neglected when $\epsilon\to 0$. To do so we split the integral in two parts. If the reaction term corresponded to that of the Fisher-Kolmogorov equation, $f(u) = u-u^2$, a different splitting is necessary.  One can show that the contribution of the nonlinear term always vanishes compared to the contribution of the linear term. A different case is that when the slope of the reaction term next to the cutoff vanishes \cite{panja1}, but we are not addressing that problem here.

\bigskip
\bigskip

\noindent
ii) {\bf Estimating the leading order of (\ref{nd})}. From the expressions (\ref{aw}) for $A$ and $\omega$  
we see that the leading order in both (\ref{N1}) and (\ref{D1}) are the terms proportional to $A^2/\omega$.
Hence, the leading order in (\ref{bdr}) is given by
\begin{equation}
J \equiv 4 \, \frac{2 \phi_* + \sin 2 \phi_*}{ 2 \phi_* + \sin 2 \phi_* + 4 \omega^2 (2 \phi_* - \sin 2 \phi_*)},
\label{B6}
\end{equation}
which we can write as
\begin{equation}
J = 4 \, \frac{1}{1+4\omega^2 \, X}
\label{B7}
\end{equation}
where $X=(2 \phi_* - \sin 2 \phi_*)/(2 \phi_* + \sin 2 \phi_*)$. 
Finally, we observe that $0< X < 1$, since $0 < \phi_* < \pi/2$ (in fact, $\phi_* \approx \pi/2$), 
and also that $1/(1+a) > (1-a)$ if $a>0$, to conclude that
\begin{equation}
J \ge 4 (1 - 4 \omega^2 X) \ge 4 (1-4 \omega^2) \approx 4 (1 - \frac{\pi^2}{(\ln \epsilon)^2}).
\label{B8}
\end{equation}

\end{document}